\documentclass[%
 reprint,
 amsmath,amssymb,
 aps,superscriptaddress
]{revtex4-2}

\usepackage{graphicx}
\usepackage{dcolumn}
\usepackage{bm}
\usepackage[normalem]{ulem}
\usepackage{xcolor}
\usepackage{mathrsfs}
\usepackage{inputenc}
\usepackage[version=4]{mhchem}
\usepackage[slantedGreek]{mathpazo}
\usepackage{hyperref}
\usepackage{float}
\usepackage{soul}
\begin{document}
\newcommand{\rb}[1]{\textcolor{red}{\it#1}}
\newcommand{\rbout}[1]{\textcolor{red}{\sout{#1}}}

\preprint{APS/123-QED}

\title{Study of ground state electronic structure of XH$^+$ (X : Cd, Hg and Yb) molecular ions via coupled-cluster approach}

\author{Ankush Thakur}
\email{ankush\_t@ph.iitr.ac.in}
\affiliation{Department of Physics, Indian Institute of Technology Roorkee, Roorkee-247667, India}
\author{Renu Bala}
\email{balar180@gmail.com}
\affiliation{Institute of Physics, Faculty of Physics, Astronomy and Informatics, Nicolaus Copernicus University, Grudziadzka 5, 87-100 Toru\'n, Poland}
\author{H. S. Nataraj}
\affiliation{Department of Physics, Indian Institute of Technology Roorkee, Roorkee-247667, India}

\begin{abstract}
The present work reports the spectroscopic parameters and molecular properties for the ground electronic state, $^1\Sigma^+$, of CdH$^+$, HgH$^{+}$, and YbH$^{+}$ molecular ions. We have used the state-of-the-art relativistic coupled cluster method together with the relativistic core-valence triple- and quadruple zeta quality basis sets for the calculation of structural parameters. The computed results have been extrapolated to the complete basis set limit using a two-point polynomial fit. The reliability of the results has been confirmed by their remarkable agreement with existing experimental and theoretical values. Further, we have calculated the relevant vibrational parameters by solving the vibrational Schr\"odinger equation using the potential energy curve and the permanent dipole moment curve of the electronic ground state. Subsequently, the lifetimes of the vibrational states have been determined by calculating the spontaneous and black-body radiation (BBR) induced transition rates. At room temperature, the lifetimes of the lowest ro-vibrational state ($v$ = \(0\), $J$ = \(0\)) due to BBR-induced transitions are estimated to be \(98.48\)\,s for CdH$^+$, \(204.85\)\,s for HgH$^+$, and \(1250.28\)\,s for YbH$^+$. Additionally, the rotational energies within each vibrational state are also calculated in this work. \\

\begin{description}
\item[Keywords]
{\it ab initio} calculations, potential energy curves, spectroscopic constants, dipole moment, quadrupole moment, dipole polarizability, transition dipole moment, ro-vibrational parameters, transition rates, lifetimes.
\end{description}
\end{abstract}
\maketitle
\section{Introduction}

Investigations for the variation of nature's fundamental physical constants are motivated by their potential to probe physics beyond the Standard Model~\cite{Calmet_2002,Uzan_2003,Stadnik_2016}. These variations are being studied extensively through both astrophysical data and laboratory experiments. Astrophysical extragalactic observations such as analyses of high-resolution quasar absorption spectra allow the examination of such variations over time intervals comparable to the age of the universe ($\sim$ $10^{10}$ yr)~\cite{Webb_2001,Flambaum_2007,Murphy_2008}, whereas laboratory measurements using different atomic and molecular clocks assess the variations over time scales of about a year~\cite{Roselband_2008,Flambaum_2009,Safronova_2019,Aeppli_2024}. In both research directions, theoretical work is indispensable for identifying suitable systems for investigation and also for interpreting how changes in measurable quantities reflect variations in fundamental constants~\cite{Borschevsky_2011}.\\

Modern laboratory experiments aimed at detecting variations rely on high-precision frequency measurements that are sensitive to the fine-structure constant ($\alpha$ = $e^{2}$/$\hbar$$c$) and the proton-to-electron mass ratio ($\mu$ = $m_p$/$m_e$)~\cite{Beloy_2011,Ganguly_2014}. In particular, molecular transition frequencies are useful for detecting potential variations in $\mu$, as vibrational and rotational transition frequencies are proportional to $\mu$$^{-1/2}$ and $\mu$$^{-1}$, respectively. In this context, several homo- and heteronuclear molecules have been proposed as potential candidates in the development of frequency standards~\cite{VFlambaum_2007,Kajita_2009,Kajita_2013,Schiller_2014,Kokish_2018,Kajita_2020}.\\

Kajita \textit{et al.}~\cite{Kajita_2011} proposed MH$^+$ (where M = $^{24}$Mg, $^{40}$Ca, $^{88}$Sr, $^{138}$Ba, $^{64}$Zn, $^{114}$Cd, $^{174}$Yb and $^{202}$Hg) molecular ions as candidate systems for detecting the temporal variation of $\mu$, owing to the possibility of measuring their pure vibrational transition frequencies with uncertainties as low as 10$^{-16}$. They can be easily cooled and trapped via sympathetic laser cooling with atomic ions confined together with the molecules~\cite{Hojbjerre_2009,Hansen_2014}. Among these ions, YbH$^+$ has been identified as the most promising for achieving the lowest measurement uncertainty~\cite{Kajita_2011}. However, transition frequencies are subject to systematic shifts, such as Stark- and blackbody radiation (BBR) shifts, which arise from the interaction of external electric fields with the molecule, thereby inducing couplings between nearby electronic, vibrational, and rotational states~\cite{Kokish_2018,MKajita_2020}. Accurate evaluation of these shifts requires precise knowledge of spectroscopic constants and molecular properties~\cite{Zalialiutdinov_2024}. Nevertheless, reliable literature data for the heavier MH$^+$ ions are scarce. Therefore, we have performed \textit{ab initio} quantum many-body calculations to determine the potential energy curves and vibrational-rotational parameters for the ground electronic state of the heavier MH$^+$ systems, and henceforth we call them XH$^+$ (where X = Cd, Hg and Yb) molecular ions. \\

Although experimental data are available for CdH$^+$ and HgH$^+$ ions~\cite{Huber_1979,Herzberg_1950}, theoretical studies that exist in the literature for the XH$^+$ ions are only a few. To mention, the spectroscopic constants of CdH$^+$ and HgH$^+$, including the equilibrium bond length ($R_e$), dissociation energy ($D_e$), and harmonic vibrational frequency ($\omega_e$), have been determined using the Dirac–Hartree–Fock (DHF) method~\cite{Ziegler_1981}, contracted configuration interaction (CCI) with effective core potentials~\cite{Stromberg_1989}, and the generalized valence bond–dissociation consistent configuration interaction (GVB-DCCI) method~\cite{Schilling_1987,Ohanessian_1990}. These spectroscopic parameters along with the anharmonic frequency ($\omega_e$$x_e$) have been computed for CdH and its ions using the relativistic Fock-space coupled-cluster (FSCC) method~\cite{Eliav_1998}. \textit{Ab initio} study on vibrational dipole moments for the ground electronic state of XH$^+$ ions have been reported by Abe \textit{et al.}~\cite{Abe_2010}, using complete active space second-order perturbation (CASPT2) theory with relativistic correlation-consistent atomic natural orbital basis sets. Among the XH$^+$ ions, the most recent study~\cite{Zhang_2014} has focused on the ground and low-lying excited states of the CdH$^+$ ion, wherein potential energy curves, spectroscopic constants, and transition properties between different electronic states have been examined using the multi-reference configuration interaction method with Davidson correction (MRCI+Q). However, the ground state results calculated using different theoretical methods show comparatively large discrepancies. Further progress toward high-accuracy results necessitates the use of highly correlated and relativistic methods in combination with larger basis sets. With this objective, we employ the fully-relativistic coupled-cluster approach, which captures extensive electron correlation effects together with a successive hierarchy of two optimized basis sets, enabling straightforward extrapolation to the complete basis set limit.\\

In the present work, we have calculated spectroscopic constants: $R_{e}$, $D_e$, $\omega_{e}$, anharmonic frequencies ($\omega_{e}$$x_{e}$ and $\omega_{e}$$y_{e}$), and rotational constants ($B_e$ and $\alpha_e$), together with molecular properties: permanent electric dipole moments (PDMs) described by $\mu_e$, components of static electric dipole polarizability ($\alpha_\parallel$ and $\alpha_\perp$), and quadrupole moments (QMs) for XH$^+$ ions. The plots corresponding to the variation of these properties as a function of the internuclear distance have also been obtained. These calculations are carried out at DHF, coupled-cluster singles and doubles (CCSD), and perturbative triples (CCSD(T)) levels of theory. Moreover, we have extended our study to the calculation of vibrational parameters such as wavefunctions, energy levels ($E_v$), vibrationally coupled rotational constants ($B_v$), transition dipole moments (TDMs) between different vibrational states, spontaneous and BBR-induced transition rates, and lifetimes of the ro-vibrational states of the considered ions. This information lays the groundwork necessary for precise spectroscopic experiments aimed at detecting variations in $\mu$.\\

The paper is organized into three subsequent sections: the introduction is followed by a detailed description of the theory and the calculations in Section~\ref{section2}, a brief discussion of the results in Section~\ref{section3}, and finally, the summary of the current work in Section~\ref{section4}.

\section{Theory and Calculational details}\label{section2}

Relativistic electronic energy calculations at DHF, CCSD, and CCSD(T) levels of theory are performed using the DIRAC23~\cite{DIRAC} software suite. We used a four-component wavefunction, which is expanded using distinct basis sets for the large and small components. To prevent variational collapse into the negative energy continuum, the kinetic balance condition is imposed on the small components of wavefunction. The Dirac-Coulomb Hamiltonian is employed in conjunction with the Visscher approximation~\cite{Visscher1997}, wherein the contribution from the (SS$\vert$SS) integrals is taken in an approximate manner. The relativistic Dyall core-valence triple-zeta (dyall.cv3z) and quadruple-zeta (dyall.cv4z) basis sets~\cite{Dyall_2007,Dyall_2004,Gomes_2010}, along with a Gaussian nuclear charge distribution, and $C_{2v}$ double group symmetry, have been utilized for these calculations. Table~\ref{table-I} provides the details of the basis functions used for the atoms constituting the diatomic molecules examined in this study. The origin of the coordinate system is fixed at the center-of-mass of the molecular ion. The molecular orbitals with energies in the range from \(-6\) $E_h$ to \(12\) $E_h$ are included in the correlation calculations to manage the computational cost. The resulting correlation space is still notably large, as indicated in Table~\ref{table-II}, which provides details of the number of active electrons and virtual orbitals incorporated into the correlation calculations. Further, the electronic energies are computed at internuclear distances starting from \(1.5\) {\AA} and extending upto \(5\) {\AA} for CdH$^+$, \(6\) {\AA} for HgH$^+$, and \(7\) {\AA} for YbH$^+$. Also, a small step size of \(0.001\) {\AA} is  considered in the proximity of equilibrium point. The dissociation energies are then obtained from the difference between the energy at the equilibrium bond length and that at the largest internuclear distance for each molecular ion. Furthermore, the VIBROT program in the MOLCAS package~\cite{OPENMOLCAS} is used to calculate spectroscopic constants from the potential energy curves (PECs) via cubic spline fitting. \\

We have employed the finite-field approach, as in Refs.~\cite{Maroulis_1998,Thakur_2024}, to calculate the first- and second-order perturbed energies of the ground states of XH$^+$ molecular ions under an external electric field. The strength of the electric field treated as perturbation is set in the range of \(-1\) $\times$ $10^{-4}$ to \(1\) $\times$ $10^{-4}$ au. From the energy derivatives with respect to the perturbative electric field, we have determined the values of PDMs and static electric dipole polarizabilities for a range of internuclear distances.\\

\begin{table}[ht]
    \centering
    \caption{\label{table-I}Details of the basis functions.}
    \begin{tabular}{c c c c}
    \hline\hline
       Atom & &Basis & Basis functions \\
       \hline
       H  &&3z & 9s, 2p, 1d \\
          && 4z & 11s, 3p, 2d, 1f \\
          \hline
       Cd & &3z & 28s, 20p, 13d, 5f, 3g \\
         & &4z & 33s, 25p, 17d, 7f, 5g, 3h \\
         \hline
       Hg && 3z & 30s, 24p, 15d, 11f, 4g, 1h \\
          && 4z & 34s, 30p, 19d, 13f, 7g, 4h, 1i \\
          \hline
       Yb && 3z & 30s, 24p, 16d, 11f, 3g, 2h \\
          && 4z & 35s, 30p, 19d, 13f, 5g, 4h, 2i \\
       \hline\hline
    \end{tabular}
    \label{tab:my_label}
\end{table}

Considering the internuclear axis of the molecular systems to be aligned along the z-axis, we have calculated the parallel (\( \alpha_\parallel \equiv \alpha_{zz}\)) and perpendicular components (\(\alpha_{xx}\; \&\; \alpha_{yy}\)) of dipole polarizability. From these, we have obtained the average, \(\bar{\alpha}\), and the anisotropic polarizability, \(\gamma\), as
\begin{equation}
    \bar{\alpha}=(\alpha_\parallel + 2\alpha_\perp)/3,
\end{equation}
\begin{equation}
    \gamma=\alpha_\parallel - \alpha_\perp.
\end{equation}
The components of the traceless quadrupole moment (QM) tensor, $\Theta_{mn}$, can be defined as~\cite{Buckingham_1959},
\begin{eqnarray}
\Theta_{mn}=\frac{1}{2}\,\sum_{i}e_i(3r_{i_m}\,r_{i_n}\,-\,r_{i}^2\,\delta_{mn}),
\end{eqnarray}
where $m$ and $n$ represent the Cartesian components, while the summation index $i$ spans over the total number of charged particles in the system. The $zz$-component of the traceless QM tensor, \emph{viz}. $\Theta_{zz}$, is related to the other diagonal components ($\Theta_{xx}$\; \&\; $\Theta_{yy}$) by the equation, 
\begin{eqnarray}
\Theta_{zz}\,=-(\,\Theta_{xx}\,+\,\Theta_{yy}).
\end{eqnarray}
For linear molecules, the components $\Theta_{xx}$ and $\Theta_{yy}$ are equal.\\
 \begin{table}[ht]
    \centering
    \caption{\label{table-II}Details of the number of active electrons and virtual orbitals for \ce{XH^+} (X: Cd, Hg and Yb) molecular systems in different basis sets.}
    \begin{tabular}{c c c c c c c}
    \hline\hline
    Molecule  & &Basis && Active electrons&& Virtual orbitals \\
    \hline
    \ce{CdH^+}  & &3z & &20 & &89 \\
                & &4z & &20&& 151 \\
    
    \hline           
    \ce{HgH^+}   & &3z & &34 && 89 \\
                 && 4z & &34 && 152 \\
    \hline
    \ce{YbH^+}  && 3z & &24 && 117 \\
            & &4z & &24 && 185 \\
    \hline\hline
    \end{tabular}
    \label{tab:my_label}
\end{table}

To overcome the incompleteness of the employed basis sets, it is necessary to extrapolate the results to the complete basis set (CBS) limit. Therefore, by utilizing energies calculated with the triple- and quadruple-zeta basis sets, we have extrapolated the energies for all molecules considered in this work to the CBS limit using the following relation~\cite{Helgaker_1997}:
\begin{equation}
    E_{n} = E_{\text{CBS}} + \frac{A}{n^3},
\end{equation}
where $A$ is a fitting parameter, $n$ is the cardinal number of the basis set, \emph{viz}. \(3\) for 3z and \(4\) for 4z, $E_{n}$ represents the energy calculated with the basis set characterized by the cardinal number $n$, and $E_{\text{CBS}}$ denotes the value of the energy at the CBS limit.\\

The vibrational Schr\"odinger equation has been solved using PECs and PDM curves computed at the CCSD(T) level in the CBS limit to obtain the vibrational spectroscopic parameters. These parameters, which include vibrational wavefunctions, vibrational-state energies ($E_v$), transition dipole moments (TDMs) between vibrational levels, and vibrationally coupled rotational constants ($B_v$), are obtained using the LEVEL program~\cite{LEVEL}. Further, the relative energy differences and TDM values between vibrational states are used to calculate the spontaneous and BBR-induced transition rates at the surrounding temperature, \textit{viz.} ($\mathrm{T}$ = \(300\) $\mathrm{K}$) as~\cite{Kotochigova_2005},

\begin{eqnarray}{}  
\Gamma_{v, J}^{spon}\,=\,\sum\limits_{v^{''}, J^{''}}\Gamma^{emis}(v, J\,\rightarrow\,v^{''}, J^{''}) 
 \end{eqnarray}
and
 \begin{eqnarray}{} 
 \Gamma_{v, J}^{BBR}\,&=& \,\sum\limits_{v^{''}, J^{''}}\bar{n}(\omega)\,\Gamma^{emis}(v, J\,\rightarrow\,v^{''}, J^{''})\nonumber\\
 &+&\sum\limits_{v^{'}, J^{'}}\bar{n}(\omega)\,\Gamma^{abs}(v, J\,\rightarrow\,v^{'}, J^{'}),
  \end{eqnarray}
respectively. Here the indices ($v^{''}, J^{''}$) and ($v^{'}, J^{'}$) denote the ro-vibrational levels within the same electronic state whose energies are lower and higher than that of the ($v, J$) level, respectively. The average number of photons $\bar{n}(\omega)$ at frequency $\omega$ is given by the relation, 
  \begin{eqnarray}{}
  \bar{n}(\omega)\,=\,\frac{1}{e^{(\hbar\omega/k_{B}T)}-1}\,,
   \end{eqnarray}
where $\hbar\omega\,=\,\arrowvert E_{v, J}-E_{\widetilde{v}, \widetilde{J}}\arrowvert$ is the energy difference between the two ro-vibrational levels involved, while $k_B$ is the Boltzmann constant. The notation ($\widetilde v, \widetilde J$) represents the ro-vibrational level with lower energy, i.e., ($v^{''}, J^{''}$) in the case of emission, while it corresponds to the higher energy level ($v^{'}, J^{'}$) for absorption.\\

The emission and absorption rates are then calculated using the equation,
\begin{eqnarray}{}
 \Gamma^{emis/abs} [(v, J)\,\rightarrow\,(\widetilde v, \widetilde J)]\,=\nonumber\\ \,\frac{8\pi}{3\epsilon_0}\frac{\omega^3}{h c^3}\, [TDM_{(v, J)\rightarrow (\widetilde v, \widetilde J)}]^2.
  \end{eqnarray}
  
Finally, the total lifetime ($\tau$) of the ro-vibrational state is obtained as,
\begin{eqnarray}{}
\tau\,=\,\frac{1}{\Gamma^{total}},
\end{eqnarray}
where $\Gamma^{total}(\,=\,\Gamma^{spon}\,+\,\Gamma^{BBR}$) is the sum of spontaneous and BBR-induced transition rates. \\

\section{Results and discussion}\label{section3}
\begin{table*}[htbp]
\begin{ruledtabular}
\begin{center}
\caption{\label{table-III}
Spectroscopic constants for the ground electronic states of the XH$^+$ molecular ions computed in this work, along with available literature values. The results presented in bold fonts are our recommended values.}
\begin{tabular}{cccccccccc}
 Basis & Method & $R_e$ & $D_e$ & $B_e$  & $\alpha_e$  & $\omega_e$ & $\omega_e x_e$ &$\omega_e y_e$ \\
& &  (\AA) & (eV) & (cm$^{-1}$) & (cm$^{-1}$) & (cm$^{-1}$) & (cm$^{-1}$)&  (cm$^{-1}$) &\\
\hline
\multicolumn{10}{c}{\textbf{CdH$^+$}}\\
\hline
 3z & DHF & 1.684 & 5.029 & 5.951 & 0.130 & 1883.49 & 26.01 & 0.573  \\
 & CCSD & 1.664 & 2.719 & 6.097 & 0.170 & 1809.55 & 33.56 & 0.195  \\
 & CCSD(T) & 1.666 & 1.693 & 6.077 & 0.183 & 1771.42 & 36.31 & 0.136  \\
 \hline
 4z & DHF & 1.684 & 5.024 & 5.952 & 0.130 & 1883.01 & 26.03 & 0.567  \\
 & CCSD & 1.664 & 2.750 & 6.101 & 0.200 & 1792.25 & 28.15 & 0.093  \\
 & CCSD(T) & 1.666 & 1.715 & 6.081 & 0.210 & 1753.30 & 38.62 & 0.393  \\
 \hline
 \textbf{CBS} & DHF & 1.684 & 5.020 & 5.953 & 0.130 & 1882.66 & 26.03 & 0.560 \\
     & CCSD & 1.664 & 2.773 & 6.101 & 0.212 & 1778.64 & 24.58 & 0.020 \\
     & \textbf{CCSD(T)} & \textbf{1.666} & \textbf{1.732} & \textbf{6.082} & \textbf{0.220} & \textbf{1739.05} & \textbf{37.57} & \textbf{0.646} \\
\hline
& Expt.~\cite{Huber_1979} & 1.6672 & 2.18 & 6.071 & 0.19 & 1772.5 & 35.40 & -  \\
& MRCI+Q~\cite{Zhang_2014} & 1.666 & 2.11 & 6.078 & - & 1789.06 & 42.59 & -  \\
& CASPT2~\cite{Abe_2010} & 1.652 & 2.1767 & 6.232 & 0.20 & 1819.6 & - & -  \\
& DHF~\cite{Pyykko_1979} & 1.825 & 1.469 &-&-&-&-&- \\
& DHF~\cite{Ziegler_1981} & 1.74 & 2.08 &-&-& 1669 &-&- \\
& CCI~\cite{Stromberg_1989}$\footnote{CCI: Contracted configuration interaction}$ & 1.70 & 1.86 &-&-& 1690 &-&- \\
& GVB-DCCI~\cite{Schilling_1987}$\footnote{GVB-DCCI: Generalized valence bond-dissociation consistent configuration interaction}$ & 1.709 & 1.93 &-&-& 1696 &-&- \\
& FSCC~\cite{Eliav_1998}$\footnote{FSCC: Fock-space coupled-cluster}$ & 1.709 & 1.908 &-&-& 1672 & 35.3 & - \\
 \hline
 \multicolumn{10}{c}{\textbf{HgH$^+$}}\\
\hline
 3z & DHF & 1.616 & 5.218 & 6.438 & 0.159 & 2069.40 & 32.64 & 0.858 \\
 & CCSD & 1.588 & 3.486 & 6.666 & 0.193 & 2086.74 & 41.11 & 0.411  \\
 & CCSD(T) & 1.589 & 2.053 & 6.656 & 0.202 & 2063.15 & 43.25 & 0.215 \\
 \hline
 4z & DHF & 1.615 & 5.223 & 6.444 & 0.159 & 2069.30 & 32.52 & 0.844  \\
 & CCSD & 1.579 & 3.593 & 6.737 & 0.177 & 2089.63 & 32.50 & 0.256  \\
 & CCSD(T) & 1.580 & 2.163 & 6.730 & 0.185 & 2067.56 & 34.20 & 0.516  \\
 \hline
  \textbf{CBS} & DHF & 1.615 & 5.227 & 6.448 & 0.159 & 2069.20 & 32.43 & 0.832 \\
      & CCSD & 1.572 & 3.671 & 6.793 & 0.169 & 2090.35 & 26.50 & 0.152 \\
      & \textbf{CCSD(T)} & \textbf{1.573} & \textbf{2.244} & \textbf{6.787} & \textbf{0.176} & \textbf{2069.31} & \textbf{27.93} & \textbf{0.797} \\
 \hline
 & Expt.~\cite{Herzberg_1950} & 1.594 & - & 6.613 & 0.206 & 2034 & - & - \\
 & CASPT2~\cite{Abe_2010} & 1.594 & 2.6626 & 6.669 & 0.224 & 2021.50 & - & - \\
 & DHF~\cite{Pyykko_1979} & 1.808 & 1.197 &-&-&-&-&-\\
 & DHF~\cite{Ziegler_1981} & 1.64 & 2.69 &-&-& 2156 &-&- \\
 & CCI~\cite{Stromberg_1989}$\footnotemark[1]$ & 1.59 & 2.52 &-&-& 2050 &-&- \\
 & GVB-DCCI~\cite{Ohanessian_1990}$\footnotemark[2]$ & 1.627 & 2.2246 &-&-& 1888 & - & - \\
 & KRCISD~\cite{Bala_2020}$\footnote{KRCISD: Kramers-restricted
configuration interaction with single and double excitations}$ & 1.588 & 4.856 & 6.682 & 0.183 & 2090.11 & 31.68 & - \\
  \hline
 \multicolumn{10}{c}{\textbf{YbH$^+$}}\\
\hline
 3z & DHF & 1.992 & 5.462 & 4.239 & 0.070 & 1496.09 & 15.78 & 0.251 \\
 & CCSD & 1.947 & 2.441 & 4.436 & 0.086 & 1491.66 & 18.39 & 1.273 \\
 & CCSD(T) & 1.949 & 1.525 & 4.379 & 0.083 & 1488.85 & 23.01 & 2.005 \\
 \hline
 4z & DHF & 1.992 & 5.459 & 4.240 & 0.070 & 1495.84 & 15.77 & 0.250 \\
 & CCSD & 1.941 & 2.484 & 4.472 & 0.146 & 1534.51 & 10.07 & 2.054 \\
 & CCSD(T) & 1.944 & 1.532 & 4.456 & 0.153 & 1527.91 & 13.12 & 2.120 \\
 \hline
  \textbf{CBS} & DHF & 1.992 & 5.456 & 4.240 & 0.070 & 1495.65 & 15.76 & 0.247 \\
      & CCSD & 1.937 & 2.515 & 4.495 & 0.181 & 1567.70 & 4.02 & 3.040 \\
      & \textbf{CCSD(T)} & \textbf{1.941} & \textbf{1.538} & \textbf{4.508} & \textbf{0.244} & \textbf{1618.95} & \textbf{5.92} & \textbf{2.419} \\
 \hline
 & CASPT2~\cite{Abe_2010} & 1.928 & 1.9735 & 4.560 & 0.098 & 1492.6 & - & - \\
 & DHF~\cite{Pyykko_1979} & 2.017 & 3.265 &-&-&-&-&- \\
\end{tabular}
\end{center}
\end{ruledtabular}
\end{table*}

\subsection{Spectroscopic constants}

\begin{figure}[htbp]
    \includegraphics[width=1.03\linewidth]{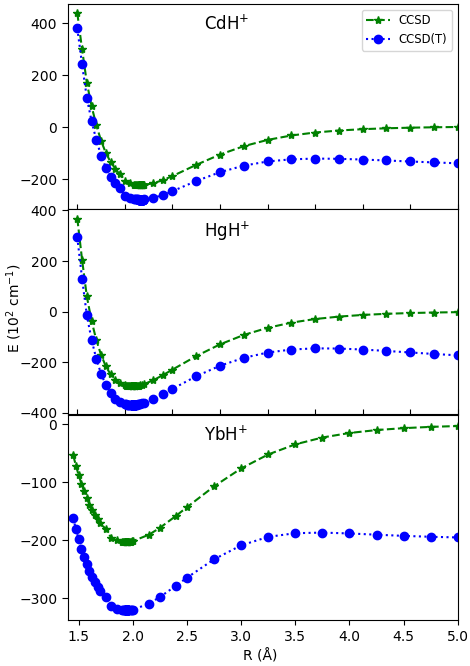}
    \caption{\label{fig:FIG1} PECs of XH$^+$ molecular ions computed at CBS limit using CCSD and CCSD(T) levels of theory, with respect to the dissociation energy at the CCSD level.}
\end{figure}

Fig.~\ref{fig:FIG1} presents the PECs calculated at the CBS limit for all XH$^+$ molecular ions using different correlation levels of theory. All PECs exhibit smooth behavior with well-defined minima. From the PECs, spectroscopic constants have been extracted and tabulated in Table~\ref{table-III}, together with the results available in the literature. The values at the CBS limit shown in bold font in the same table are our recommended results. \\

For the CdH$^+$ ion, the computed diatomic constants agree closely with experimental measurements~\cite{Huber_1979}, with individual differences of \(0.001\) {\AA} for $R_e$, \(0.448\) eV for $D_e$, \(0.011\) cm$^{-1}$ for $B_e$, \(0.03\) cm$^{-1}$ for $\alpha_e$, \(33.45\) cm$^{-1}$ for $\omega_e$, and \(2.17\) cm$^{-1}$ for $\omega_ex_e$. A comparison with Ref.~\cite{Zhang_2014} shows that the $B_e$ and $\omega_e$ values at the MRCI+Q level differ from ours by \(0.1\)\% and \(2.7\)\%, respectively. While $R_e$ is identical in both studies, $D_e$ and $\omega_ex_e$ deviate by \(0.37\) cm$^{-1}$ and \(5.02\) cm$^{-1}$, respectively. Besides the different computational methods employed, these discrepancies could also be ascribed to the smaller basis set used for Cd in their calculations, as against the larger 4z basis set utilized in the present work. Our diatomic constants indicate a difference of utmost \(4\)\% in $R_e$, $B_e$, and $\omega_e$ values compared with the CASPT2 results reported by Abe \textit{et al.}~\cite{Abe_2010}, likely owing to the ability of CC method to capture more correlation effects in our calculations. The other available results in the literature  using CCI~\cite{Stromberg_1989}, GVB-DCCI~\cite{Schilling_1987}, and FSCC~\cite{Eliav_1998} methods show a maximum difference of \(2.5\)\% for $R_e$, \(10\)\% for $D_e$, and \(4\)\% for $\omega_e$ relative to our recommended values. \\

The difference between our calculated constants and the experimental data~\cite{Herzberg_1950} for the HgH$^+$ ion remains within \(2.6\)\%, reflecting good agreement. For the YbH$^+$ ion, no experimental results are available yet, and even the theoretical calculations are rather scarce compared to other molecular ions. In comparison with Ref.~\cite{Abe_2010}, the maximum differences in the spectroscopic constants ($R_e$, $B_e$, and $\omega_e$) are found to be \(2\)\% for HgH$^+$ and \(8\)\% for YbH$^+$. Meanwhile, the $D_e$ values differ by \(0.42\) eV and \(0.44\) eV for HgH$^+$ and YbH$^+$ molecular ions, respectively. The observed differences may stem from distinct choices of active spaces for correlation, basis sets, and computational methodologies. In the case of HgH$^+$ ion, the differences between our reported values and those available in the literature~\cite{Pyykko_1979,Ziegler_1981,Stromberg_1989,Ohanessian_1990} span \(0.017\)-\(0.235\) {\AA} for $R_e$, \(0.019\)-\(1.047\) eV for $D_e$, and \(19.31\)-\(181.31\) cm$^{-1}$ for $\omega_e$. However, when compared to the Kramers-restricted configuration interaction with single and double excitations (KRCISD) results~\cite{Bala_2020}, the maximum difference is about \(3.8\)\% for all constants except $D_e$ and $\omega_e$. The accuracy of these findings can be further validated by future experimental measurements, as the precision of the available data from earlier experiments is not known. \\

Figure~\ref{fig:FIG1} reveals that YbH$^+$ exhibits the shallowest potential well among the XH$^+$ ions, characterized by the lowest dissociation energy. The incorporation of electron correlation effects lowers both $R_e$ and $D_e$ for all molecular ions. The maximum contributions of perturbative triples to the CCSD values are approximately \(0.2\)\% for $R_e$, \(0.3\)\% for $B_e$, and \(3.3\)\% for $\omega_e$. In particular, $D_e$ seems to be highly sensitive to higher-order correlation effects, which contribute nearly \(38\)\%. We have observed the minimum value of $R_e$ for HgH$^+$, which corresponds to the maximum $B_e$ and $\omega_e$. For each individual molecular ion, the $\alpha_e$ values obtained at the CCSD and CCSD(T) levels are of the same order, and the same is true for $\omega_ex_e$. \\

We have also investigated the dissociative behavior of XH$^+$ molecular ions by comparing the sum of the atomic energies of X$^+$ and H atoms with the molecular energy at the dissociation limit, both calculated in Hartrees at the highest level of correlation (CCSD(T)). The relative percentage difference between them does not exceed \(0.0005\)\%. This confirms that XH$^+$ ions dissociate into X$^+$ and H atoms at the dissociation limit. The nature of the asymptotic molecular states identified in this work is consistent with those published in the literature~\cite{Schilling_1987,Ohanessian_1990,Abe_2010,Zhang_2014}. Further, we have observed a slight decrease in the CCSD(T) energy values beyond an internuclear distance of \(3.5\) {\AA}. However, this does not significantly affect the $D_e$ results, as confirmed by the negligible energy difference between the sum of atomic (X$^+$ and H) and molecular energies at the dissociation limit. \\

\subsection{Molecular Properties}

\begin{table*}[htbp]
\caption{\label{table-V}
Magnitude of molecular properties (electric dipole ($\mu_e$) and electric quadrupole moments ($\Theta_{zz}$), and components of dipole polarizabilities ($\alpha_\parallel, \alpha_\perp$, $\bar{\alpha}$ and $\gamma$)) at the complete basis set limit (CBS) for the ground state of XH$^+$ molecules. All the properties are given in atomic units (au). The results presented in bold fonts are our recommended values.}
\begin{ruledtabular}
\begin{tabular}{cccccccc}
Molecule & Method &$\mu_e$  & $\Theta_{zz}$ & $\alpha_\parallel$ & $\alpha_\perp$ & $\bar{\alpha}$ & $\gamma$\\
\hline
\textbf{CdH$^+$} & DHF & 1.032 & 6.668 & 27.660 & 19.660 & 22.327 & 8.000 \\
        & CCSD & 0.642 & 6.736 & 30.450 & 20.060 & 23.523 & 10.390 \\
        & \textbf{CCSD(T)} & \textbf{0.568} & \textbf{6.758} & \textbf{31.430} & \textbf{20.260} & \textbf{23.983} & \textbf{11.170} \\
        & MRCI+Q~\cite{Zhang_2014}\footnote{The PDM has been extracted from the graph given in Refs.~\cite{Abe_2010,Zhang_2014}.} & 0.510 & - & - & - & - & - \\
        & CASPT2~\cite{Abe_2010}$\footnotemark[1]$ & 0.775 & - & - & - & - & - \\
        \hline
\textbf{HgH$^+$} & DHF & 0.523 & 8.908 & 27.360 & 20.620 & 22.867 & 6.740 \\
        & CCSD & 0.191 & 8.974 & 27.160 & 20.110 & 22.460 & 7.050 \\
        & \textbf{CCSD(T)} & \textbf{0.135} & \textbf{8.999} & \textbf{27.530} & \textbf{20.080} & \textbf{22.563} & \textbf{7.450} \\
        & CASPT2~\cite{Abe_2010}$\footnotemark[1]$ & 0.346 & - & - & - & - & - \\
        & KRCISD~\cite{Bala_2020} & 0.342 & - & - & - & - & - \\
        \hline
\textbf{YbH$^+$} & DHF & 2.111 & 7.520 & 38.860 & 22.020 & 27.633 & 16.840 \\
        & CCSD & 1.711 & 7.620 & 47.520 & 23.770 & 31.687 & 23.750 \\
        & \textbf{CCSD(T)} & \textbf{1.661} & \textbf{7.649} & \textbf{49.900} & \textbf{23.300} & \textbf{32.167} & \textbf{26.600} \\
        & CASPT2~\cite{Abe_2010}$\footnotemark[1]$ & 1.809 & - & - & - & - & - \\
\end{tabular}
\end{ruledtabular}
\end{table*}

A range of molecular properties, including PDMs, quadrupole moments ($\Theta_{zz}$), and components of the static electric dipole polarizabilities ($\alpha_\parallel, \alpha_\perp$, $\bar{\alpha}$ and $\gamma$), have been computed for the XH$^+$ ions. The results evaluated at the equilibrium bond length for different levels of theory at the CBS limit are tabulated in Table~\ref{table-V}.\\

Although, the PDM results are available in the literature, the results for other properties are reported in this work for the first time, to the best of our knowledge. The present systems are ionic molecules, and the value of the PDM depends on the choice of the coordinate origin. We have taken the origin as the center-of-mass of the molecular ions. At the CCSD(T) level, our calculated PDM values (in atomic units (au)) are lower by \(0.20\), \(0.21\), and \(0.15\) for CdH$^+$, HgH$^+$, and YbH$^+$, respectively, compared to the results reported at the CASPT2 level~\cite{Abe_2010}. For the CdH$^+$ ion, Zhang \textit{et al.}~\cite{Zhang_2014} have reported the PDM using MRCI+Q theory. Their value compares well with our recommended value, with a difference of about \(0.06\) au at the CCSD(T) level. These differences between our results and those reported in the literature could be attributed to the use of different methods, basis sets, and configurational space. It has to be noted, however, that the reported values of PDMs in Refs.~\cite{Zhang_2014,Abe_2010} are extracted from their PDM curves at the equilibrium bond length. The largest PDM value is observed for YbH$^+$, likely due to its high electronegativity difference and the longest $R_e$ among the molecular ions. Compared to the DHF result, the total electron correlation contribution to $\Theta_{zz}$ at the CCSD(T) level is \(1.4\)\% for CdH$^+$, \(1\)\% for HgH$^+$, and \(1.7\)\% for YbH$^+$. Further, YbH$^+$ exhibits the most pronounced correlation effects in $\bar{\alpha}$ and $\gamma$, with fractional percentage deviations (relative to DHF) of approximately 16\% and 58\%, respectively. We also observe that the leading-order triple excitations contribute (\(0.3\)\%, \(0.3\)\%, \(0.4\)\%) over the CCSD values of ($\Theta_{zz}$, $\bar{\alpha}$, $\gamma$) for CdH$^+$, (\(2\)\%, \(0.5\)\%, \(1.5\)\%) for HgH$^+$, and (\(7.5\)\%, \(5.7\)\%, \(12\)\%) for YbH$^+$. \\

\begin{figure}[]
    \includegraphics[width=1.03\linewidth]{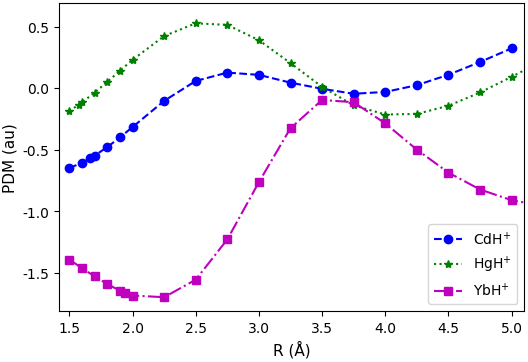}
    \caption{\label{fig:FIG2}CBS values of PDM as a function of internuclear distance for XH$^+$ molecular ions, computed at CCSD(T) level of theory.}
\end{figure}

Fig.~\ref{fig:FIG2} shows the behavior of the PDMs of XH$^+$ molecular ions as a function of internuclear distance (R) in the range of \(1.5\)–\(5\) {\AA}, calculated at the CCSD(T)/CBS level. The PDM curves of CdH$^+$ and HgH$^+$ exhibit relatively similar trends: the values initially increase to a maximum, then decrease up to about R = \(4\) {\AA}, and subsequently rise again at larger R. Such an increase at large R values is typically observed for heteronuclear molecular ions~\cite{ElOualhazi_2016, Habli_2016}. Moreover, the sign of the PDM changes with increasing bond length. In contrast, the PDM curve of YbH$^+$ remains entirely negative throughout the bond length range. Notably, at short internuclear distances, the observed variations in PDMs arise from the strong sensitivity of the molecular system to changes in its charge distribution, which reflects the nature of interatomic interactions within the molecule. \\

\begin{figure}[]
    \includegraphics[width=1.03\linewidth]{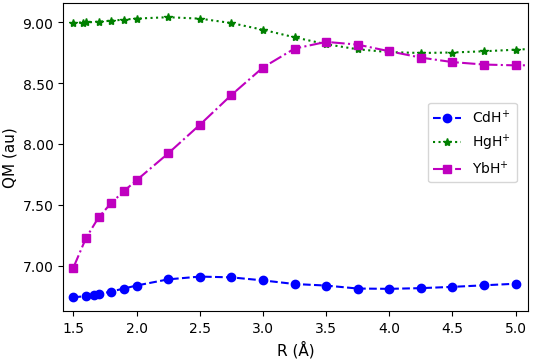}
    \caption{\label{fig:FIG3}CBS values of QM as a function of internuclear distance for XH$^+$ molecular ions, computed at CCSD(T) level of theory.}
\end{figure}

\begin{figure}[]
    \includegraphics[width=1.03\linewidth]{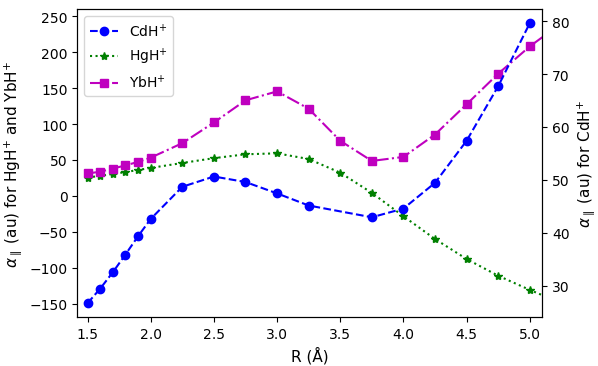}
    \caption{\label{fig:FIG4}CBS values of parallel component of static dipole polarizability ($\alpha_\parallel$) as a function of internuclear distance for XH$^+$ molecular ions, computed at CCSD(T) level of theory.}
\end{figure}

The R-variation of QMs for all molecular ions is shown in Fig.~\ref{fig:FIG3}. For CdH$^+$ and HgH$^+$ ions, the curves reach a maximum at R = \(2.5\) {\AA}, after which they decrease and gradually approach nearly constant values at larger separations. However, YbH$^+$ shows a significant increase in the QM as R increases from \(1.5\) to \(3.5\) {\AA}, beyond which it follows a trend similar to that of the other molecular ions. Furthermore, the dependence of the parallel component of static electric dipole polarizability ($\alpha_\parallel$) on R is presented in Fig.~\ref{fig:FIG4}. YbH$^+$ is found to have the largest polarizability over the entire range. An understanding of static polarizability and its dependence on R is required for calculating thermal Stark shifts in molecules~\cite{Zalialiutdinov_2024}. \\

As we have used large basis sets (3z and 4z) and extrapolated the energies to the CBS limit, the basis set superposition error (BSSE) in the energy calculations is expected to be minimal~\cite{Miliordos_2015}. Since the properties are calculated as derivatives of the CBS energy, BSSE should not have a notable effect on our computed results. However, errors may arise from the exclusion of additional active electrons and virtual orbitals in the restricted configurational space considered here. To pursue this further, we have expanded the configurational space for the heaviest ion, HgH$^+$, from \(-6\) $E_h$ to \(12\) $E_h$ to an extended range of \(-10\) $E_h$ to \(30\) $E_h$. This relaxation has included an additional \(10\) active electrons and \(37\) virtual orbitals, at the cost of significant computational expense. As a result, we have observed changes of \(3\)\%, \(6\)\%, and \(0.3\)\% in the values of $D_e$, PDM, and $\alpha_\parallel$, respectively. Based on these findings, we anticipate that the maximum error in our final results for the spectroscopic constants and molecular properties may not exceed \(6\)\%. \\

\subsection{Vibrational parameters}

\begin{figure}[htbp]
    \includegraphics[width=1.03\linewidth]{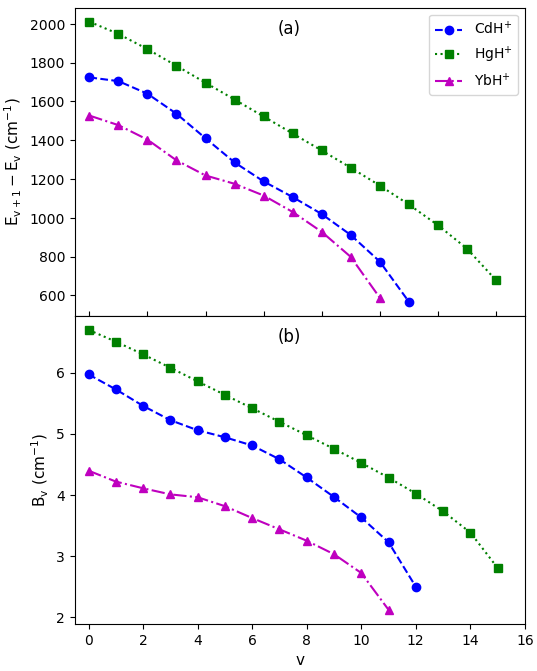}
    \caption{\label{fig:FIG5}The figure shows (a) energy spacings between adjacent vibrational levels and (b) vibrationally coupled rotational constants for XH$^+$ molecular ions at CCSD(T) level of theory.}
\end{figure}

On solving the vibrational Schr\"odinger equation using the PECs and PDM curves obtained at the CCSD(T)/CBS level, we have calculated all the vibrational parameters. The number of vibrational states is found to be \(12\) for CdH$^+$, \(15\) for HgH$^+$, and \(11\) for YbH$^+$. The relatively deeper potential well of HgH$^+$ results in the largest number of bound vibrational levels. Fig.~\ref{fig:FIG5}(a) and Fig.~\ref{fig:FIG5}(b) present the energy spacing between successive vibrational levels ($E_{v+1}-E_v$) and vibrationally coupled rotational constants ($B_v$), respectively, as a function of vibrational quantum number ($v$). The trends exhibit a gradual decrease, reflecting the anharmonicity of the potential well. It is also evident that the vibrational levels are non-equidistant and tend to converge as $v$ increases. It has been shown in the literature that variations in the proton-to-electron mass ratio are more sensitive to the deeply bound vibrational levels at intermediate internuclear distances~\cite{Zelevinsky_2008,Kotochigova_2009}. \\

\begin{figure}[]
    \includegraphics[width=1.03\linewidth]{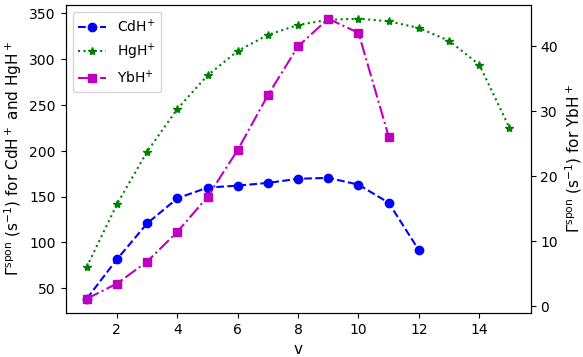}
    \caption{\label{fig:FIG6}The figure shows spontaneous transition rates for the vibrational levels of XH$^+$ molecular ions.}
\end{figure}

\begin{figure}[]
    \includegraphics[width=1.03\linewidth]{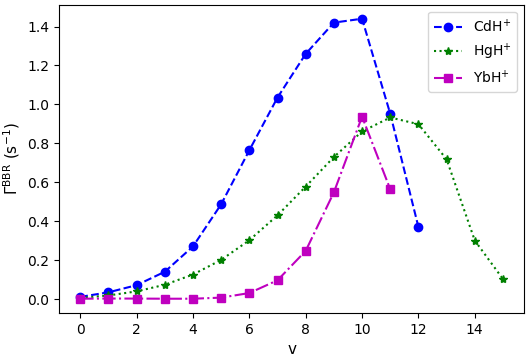}
    \caption{\label{fig:FIG7}The figure shows BBR-induced transition rates, at room temperature (T = \(300\) K), for the vibrational levels of XH$^+$ molecular ions.}
\end{figure}

\begin{figure}[]
    \includegraphics[width=1.03\linewidth]{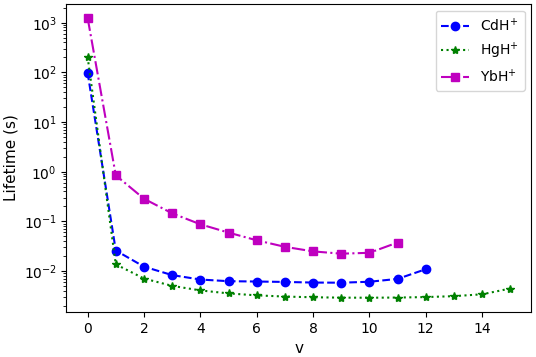}
    \caption{\label{fig:FIG8}The figure shows lifetimes, at room temperature (T = \(300\) K), for the vibrational levels of XH$^+$ molecular ions.}
\end{figure}

Using the vibrational energy spacings and TDMs between different vibrational states, we have calculated the spontaneous and BBR-induced transition rates. All calculated TDMs for vibrational state transitions are provided in Supplementary Table S1. For the BBR rate calculations, we considered a surrounding temperature of \(300\) $\mathrm{K}$. We have shown the variations in the spontaneous and BBR-induced transition rates for the vibrational states of XH$^+$ against $v$ in Fig.~\ref{fig:FIG6} and Fig.~\ref{fig:FIG7}, respectively. Further, the vibrational state lifetimes are obtained from the inverse of the total transition rates and are plotted in Fig.~\ref{fig:FIG8} with respect to $v$. We observe that the lower vibrational levels have longer lifetimes than the higher vibrational levels for all molecular ions. The lifetimes of the lowest ro-vibrational states ($v$ = \(0\), $J$ = \(0\)) at \(300\) $\mathrm{K}$ are governed exclusively by BBR-induced transitions and are evaluated to be \(98.48\)\,s, \(204.85\)\,s, and \(1250.28\)\,s for CdH$^+$, HgH$^+$, and YbH$^+$, respectively. Previous studies have likewise reported long lifetimes of the lowest ro-vibrational state for other molecular ions~\cite{Zrafi_2020,Ladjimi_2022,Zrafi_2023}. \\

As illustrated in Fig.~\ref{fig:FIG8}, the lifetimes first decrease sharply, then reach a minimum, and finally exhibit a slight increase for larger $v$. This increase in lifetime can be explained by the variation of spontaneous (Fig.~\ref{fig:FIG6}) and BBR-induced (Fig.~\ref{fig:FIG7}) transition rates. The shortest lifetimes are observed at $v$ = \(9\), \(10\), and \(9\) for CdH$^+$, HgH$^+$, and YbH$^+$, respectively, which coincide with the peak in the spontaneous transition rate and lie close to the maximum of the BBR-induced transition rate. Beyond these values, both transition rates fall as the energy spacing between higher vibrational states is rendered progressively smaller than that of the lower states. Consequently, the vibrational state lifetimes slightly increase, a trend that has also been reported for several molecules in the literature~\cite{Zrafi_2023,Gopakumar_2011,Fedorov_2017,Bala_2019,Thakur_2025}. These lifetime calculations are highly valuable for ultracold experiments, supporting the study of long-range dipole–dipole interactions and enabling precise measurements to probe potential time variations in the proton-to-electron mass ratio~\cite{Ospelkaus_2010}. \\

Furthermore, we have calculated the rotational energies of each vibrational state for all molecular ions. As anticipated, the rotational levels within higher vibrational states are more closely spaced than those in lower vibrational states. The rotational energies are listed in Supplementary Table S2. 

\section{Summary}\label{section4}
To summarize, we have performed \textit{ab initio} calculations for the electronic ground state structure of CdH$^+$, HgH$^+$, and YbH$^+$ molecular ions. To obtain precise results for the spectroscopic constants and molecular properties, we have used the four-component relativistic CC method, employing large uncontracted basis sets with extrapolation to the CBS limit. For CdH$^+$ and HgH$^+$, where experimental measurements of the spectroscopic constants are available, our results are in good agreement with the measured values. We have also compared our calculated spectroscopic constants and PDMs with theoretical results available in the literature. Further, the results of the static electric dipole polarizabilities and QMs for all the molecular ions are being reported here for the first time, to the best of our knowledge. The maximum error in our final results is estimated to be about \(6\)\% or less, resulting from the expansion of the correlation space. \\

Finally, from the PECs and PDM curves, we have determined the vibrational wavefunctions, $E_v$, TDMs between vibrational states, and $B_v$. The spontaneous and BBR-induced transition rates, and hence the lifetimes of the vibrational levels, have also been calculated. The lifetimes of the ro-vibrational ground state are found to be \(98.48\)\,s for CdH$^+$, \(204.85\)\,s for HgH$^+$, and \(1250.28\)\,s for YbH$^+$, at room temperature. In addition, the rotational energies have been calculated for each vibrational state. \\

The accurate spectroscopic constants and molecular properties are expected to be useful for future experimental and theoretical investigations focused on advancing fundamental physics and timekeeping technology. 

\begin{acknowledgments}
We would like to thank the National Supercomputing Mission (NSM) for providing computing resources of ‘PARAM Ganga’ at the Indian Institute of Technology Roorkee, implemented by C-DAC and supported by the Ministry of Electronics and Information Technology (MeitY) and Department of Science and Technology (DST), Government of India. Some of the calculations reported in this work were performed on the high-performance computing facility available in the Department of Physics, IIT Roorkee, India. We would like to acknowledge Roman Ciury\l{}o and Piotr S. \.Zuchowski of Nicolaus Copernicus University for their careful reading of the manuscript and for their valuable comments. R.B. was supported by Polish National Science Centre Project No. 2021/41/B/ST2/00681. The research is also a part of the program of the National Laboratory FAMO in Toru\'n, Poland.
\end{acknowledgments}
\bibliography{XH+_GS}

\end{document}